# Time-Reversal Symmetry and the Structure of Superconducting Order Parameter of Nearly Ferromagnetic Spin-Triplet Superconductor UTe$_2$

V.G. Yarzhemsky [1,2*], E.A.Teplyakov[1,2]

[1] Kurnakov Institute of General and Inorganic Chemistry of RAS, 31 Leninsky prospect, 119991 Moscow, Russia

[2] Moscow Institute of Physics and Technology, Dolgoprudny, 9 Institutsky lane, 141701, Moscow Region, Russia

*Corresponding author e-mail vgyar@igic.ras.ru

**Abstract**

The space-group approach to the wavefunction of a Cooper pair is extended to Shubnikov groups of type III and is applied to newly discovered nearly ferromagnetic superconductor UTe$_2$. A nonunitary triplet SOP (superconducting order parameter), manifested experimentally, was obtained for Shubnikov group $m'm'm$. The nodal structure of triplet nonunitary SOP is investigated and it was shown that in the case of Shubnikov III symmetry nodal structure of ESP (equal spin pairing) state is in some cases unstable, i.e. changes when a complex phase winding is added. The irreducible corepresentations of a Cooper pair wave function corresponding to $m'm'm$ symmetry and to all experimental data for UTe$_2$ are identified.

Triplet superconductors are characterized by complex structure of SOP (superconducting order parameter), which can be unitary or nonunitary and reveals nontrivial topological aspects [1,2]. The uranium based superconductors UPt$_3$ [3-7], URhGe, UCoGe, and UGe$_2$ [8] manifest many of these unconventional features, namely anisotropic $p$-wave pairing with line and (or) point nodes, broken time-reversal symmetry and chirality of SOP. Recently discovered uranium based superconductor UTe$_2$ [9] exhibits crucial features of a spin-triplet pairing state, i.e. an extremely large, anisotropic upper critical field $H_{c2}$, temperature-independent NMR Knight shift, and power law temperature dependence of several physical values. On the contrast to the above mentioned materials, whose SOP is of triplet OSP (opposite spin pairing) type [7,8] the SOP of UTe$_2$ is triplet ESP (equal spin pairing) and nonunitatry [9]. A large fraction of ungapped electronic states of UTe$_2$ may correspond to the nonunitary triplet state, in which two-component superconducting order parameter has two different energy gaps [9.10]. It was mentioned [9], that triplet nonunitary state is generally not expected for paramagnetic, orthorhombic systems with strong spin-orbit coupling [9]. The hallmark of this material among the other triplet superconductors is a very high magnetic field $H_{c2}$ 45 T and extreme magnetic field-boosted superconductivity at 65 T [11]. These unusual features also correspond to nonunitary pairing.

Power law temperature dependences of broader number of physical values of UTe$_2$ in superconducting state were interpreted in terms of superconducting gap structure with point nodes, and the authors pointed to the incompatibility of nonunitary nature of SOP and point group $D_{2h}$ symmetry[12]. On the basis of microwave surface impedance measurements of UTe$_2$ crystals it was concluded that UTe$_2$ may be the first example of the $3D$ chiral triplet superconductor [13]. Scanning tunneling microscopy [14] combined with existing data indicating triplet pairing [9, 10] and also suggests chiral SOP with broken time-reversal symmetry. Experiments under a pressure revealed another superconducting transition at $T_c$=3K [16].

Calculations by GGA+U methd resulted in band closing and SOP of UTe$_2$ was classified in terms of spin components $\hat{x}$, $\hat{y}$, $\hat{z}$ and continues $k$-vector, and point nodes were obtained for $D_{2h}$ symmetry [17]. Similar theoretical investigation of UTe$_2$ were performed and the authors pointed out "that previously proposed nonunitary pairing is inconsistent with the experimental implication of point nodes and therefore excluded"[18].
Nonunitary SOP was constructed for the direct product group $SO(3) \times D_{2h}$, i.e. in a weak spin orbit coupling case only [19]. However nonunitary SOP in a strong spin-orbit coupling case can't be described in terms of basis functions of one-dimensional real IRs of point group $D_{2h}$. and the search for other underlying symmetries is required. In a conventional superconductor gauge symmetry is broken, while unconventional superconductors break other symmetries as well [20, 21]. In a general case symmetry group of a Cooper pair is a direct product of point group and time-reversal [22]. Low-temperature μSR measurements on single crystals of UTe$_2$ found no evidence of long-range or local magnetic order down to 0.025 K, but indicate that the superconductivity coexists with the magnetic fluctuations [23]. Hence it follows that magnetic symmetry [24] is important for the description of a Cooper pairs in UTe$_2$.

The present work has two aims. Firstly to extend group theoretical description of a Cooper pair wavefunction on Shubnikov groups of type III, and secondly to find SOP of UTe$_2$, corresponding to its experimentally established properties, namely, $D_{2h}$ point group symmetry and nonunitary SOP with point nodes in a strong spin-orbit coupling case. These aims are achieve using space-group approach to the wavefunction of a Cooper pair [25-30], which is based on the exact space group IRs [24] for one electron wavefunctions and Mackey-Bradley theorem for pair functions [31,32].

The space group approach to the wavefunction of a Cooper pair is based on the Pauli exclusion principle and the exact symmetry of one-electron wavefunctions in crystals. The latter ones are transformed by induced representations $D^k \uparrow G$, where $D^k$ is IR of the group $H$ of the wavevector $k$ [24]. Thus the spatial part of a triplet pair belongs to antisymmetrized Kronecker square of one-electron IR $\{(D^k \uparrow G) \times (D^k \uparrow G)\}$. The decomposition of this Kronecker square consists of several stars of wave vectors for two-electron states. Since total momentum of a Cooper pairs is zero, only one term in decomposition of antisymmetrized square is considered. Its characters $\chi(P^{k-}(m))$ are constructed on the group $M=H+IH$ and are written as [25, 31, 32]:

$$\chi(P^{k-}(h)) = \chi(D^k(h))\chi(D^k(IhI)) \qquad (1)$$

$$\chi(P^{k-}(Ih)) = -\chi(D^k(IhIh)) \qquad (2)$$

Where $h \in H$. Possible representations for spatial parts of triplet pairs are obtained by the induction of the characters, defined by formulae (1) and (2), into $G$ [24-27]. Representation $P^{k-} \uparrow G$ can be easily decomposed into irreducible components making use of Frobenius reciprocity theorem [24, 25]

Odd two-electron basis function for $k$ a general point in a BZ (Brillouin zone) is written as [26]:

$$\psi_k = \varphi_k(r_1)\psi_{Ik}(r_2) - \varphi_k(r_1)\psi_{Ik}(r_2) \qquad (3)$$

The action of a left cost representatives in a left coset decomposition of the group $G$ relative to a subgroup $C_i$ results in total Coope pair basis set. In the case of $D_{2h}$ point group it is convenient to choose coset representatives $\sigma_x$, $\sigma_y$ and $C_{2z}$. This basis set is shown in Figure 1.

In the case of $D_{2h}$ group with time reversal, total pair function of triplet pair is a direct product of its spatial part by one of possible spin parts $\hat{x}$, $\hat{y}$ and $\hat{z}$. To construct total pair function one can multiply spatial part (3) by the spin part and make use of projection operator technique [26] for spin-orbital $\hat{x}\psi_k$ ($\hat{z}\psi_k$ or $\hat{y}\psi_k$). Triplet pair functions for $D_{2h}$ group and their nodal structures are presented in Table 1. Consider a nodeless spatial odd pair function on the plane [010] which transforms as $B_u$ of $C_{2h}$ on this plane. On this plane $\hat{y}$ spin transforms as $A_g$ and $\hat{x}$ and $\hat{z}$ spins transform as $B_g$. When multiplying these spin components by the spatial part of triplet pair we obtain all two odd IRs of $C_{2h}$. It is obvious that further induction will result all odd IRs of the whole group. Hence it follows that in a space-group approach Blount theorem [33] is also valid. Violations o Blount theorem are possible for non-symmorphic groups on the surface of BZ [25, 27-29]. The basis functions for $D_{2h}$ group obtained according to phenomenological approach [20] include products of spin function by the components of continuous $k$-vector [12, 17, 18]. Basis functions [12, 17] for $B_{2u}$ are similar to each other and include functions $k_x\hat{z}$ and $k_z\hat{x}$, which don't vanish simultaneously in any plane, but the first one vanishes in [100] plane and the second one vanishes in the [001] plane. The same results are seen in Table 1 for functions $\hat{z}B_{2u}$ and $\hat{x}B_{2u}$. Despite of the fact that the nodes in the two approaches coincide in this case, their fundamental physical principles are different. In phenomenological approaches [12, 17, 20, 21] nodal structures of polynomial functions, which may contain different powers of isotropic $k$-vector are analyzed. In a space-group approach $k$-vector runs over the representation domain of a BZ [24] and the basis functions in other domains, whose number equals to $|G|/2$, are obtained group theoretically (see Figure 1).

Experimental data indicate time-reversal symmetry violation in UTe$_2$ [9, 23]. Hence the initial symmetry group $\theta \times D_{2h}$ (grey Shubnikov group) is subduced to one of its subgroups. The $D_{2h}$ group is not an appropriate candidate for nonunitary case, since all its IRs are one-dimensional and real [9,12,17,18]. The black and white Shubnikov groups $m'm'm'$ and $mmm'$ also would not do, since in these group time reversal is connected with space inversion, but the operation $\theta I$ makes possible to construct OSP component of triplet pair or singlet pair only [22]. It remains just a group $m'm'm$ in which time-reversal is connected with the elements $C_{2x}$, $C_{2y}$, $\sigma_x$ and $\sigma_y$. The ICRs (irreducible corepresentations) of this Shubnikov group belong to the type ***a*** [24] and are obtained as follows. For unitary subgroup $C_{2h}$ they coincide with its IRs $A_g$, $A_u$, $B_g$ and $B_u$. They are extended on nonunitary elements with sign + or -. We will use this signs for notation of IR of $m'm'm$. When taking $\theta C_{2y}$ as a nonunitary left coset representative we obtain that the characters of four odd ICRs of $m'm'm$, namely $A_u^+$, $A_u^-$, $B_u^+$ and $B_u^-$ are identical with the characters of IRs $B_{1u}$, $A_u$, $B_{3u}$ and $B_{2u}$ of $D_{2h}$ respectively. In magnetic field spin triplet function has three components $S_0^1$ and $S_{\pm1}^1$, where subscripts are magnetic quantum numbers. We use basis functions $S_{\pm1}^1 = \hat{x} \pm i\hat{y}$ for ESP functions. In basal plane spin functions $S_{\pm1}^1 = \hat{x} \pm i\hat{y}$ belong to IR $B_g$ of $C_{2h}$, spatial part belongs to $B_u$ and total function belongs to $A_u$. The extension of $A_u$ to $m'm'm$ results in IRs $A_u^+$ and $A_u^-$. Symmetry group of vertical planes is Shubnikov group $2'/m'$. Transformation properties of spin functions on vertical symmetry planes are as follows:

$$\theta\sigma_x(\hat{x}+i\hat{y}) = \theta(\hat{x}-i\hat{y}) = \hat{x}+i\hat{y} \qquad (4)$$

$$\theta\sigma_y(\hat{x}+i\hat{y}) = \theta(-\hat{x}+i\hat{y}) = -\hat{x}-i\hat{y} \qquad (5)$$

It is clear that the replacement of $\sigma_{x(y)}$ by $C_{2x(y)}$ in these relations gives the same result. It is also obvious that transformation properties of spin-down component $\hat{x}-i\hat{y}$ are the same. However, if spin function is multiplied by complex phase $i$ the relations on the planes [100] and [010] interchange:

$$\theta\sigma_x(i\hat{x}-\hat{y}) = \theta(i\hat{x}+\hat{y}) = -i\hat{x}+\hat{y} \qquad (6)$$

$$\theta\sigma_y(i\hat{x}-\hat{y}) = \theta(-i\hat{x}-\hat{y}) = i\hat{x}-\hat{y} \qquad (7),$$

Transformation properties of total pair functions in planes are obtained as that of products of spin part by spatial part $B_u$ The pairs' wavefunctions obtained by projection operators and their nodal structures are presented in Table 2. Thus we obtained non-unitary ESP pair functions, which belong to ICRs of the subgroup of total symmetry group $\theta \times D_{2h}$, namely, Shubnikov group $m'm'm$ and the above mentioned contradiction between point group theory and nonunitary nature of SOP in UTe$_2$ is eliminated.

Nodal structures of triplet pairs, corresponding to the data of Table 2 for $A_u^+$ and $A_u^-$. are shown in Figure 2 a) – d). It is seen, that the application of Shubnikov group theory resulted in that the nodes of ICRs depend on a complex phase factor and this new result requires additional discussion. According to a space-group approach to the wavefunction of a Cooper the wave vector $k$ in runs over $1/2|G|$ of the whole BZ and the functions corresponding to other prongs are obtained group-theoretically (see Figure 1). Nevertheless inside this domain the phase of a Cooper pair wavefunction can change, resulting in a phase winding. It is seen from Figures 2 a) and 1 b), that function $A_u^+(\hat{x}+i\hat{y})$ is nodal in plane [010] and nodeless in plane [100] and that function $A_u^+(i\hat{x}-\hat{y})$ is nodal in plane [100] and in nodeless in plane [010]. Suppose that there is a phase winding $e^{i\varphi}$ accompanying the rotation of $k$ from 0 to $\pi/2$. When starting from functions $A_u^+(\hat{x}+i\hat{y})$ and $A_u^+(i\hat{x}-\hat{y})$ on [010] plane we obtain on plane [100] functions $e^{i\pi/2}A_u^+(\hat{x}+i\hat{y}) = A_u^+(i\hat{x}-\hat{y})$ and $e^{i\pi/2}A_u^+(i\hat{x}-\hat{y}) = A_u^+(-\hat{x}-i\hat{y})$ respectively. The first function is nodal in [100] plane and the second function is nodeless in this plane. Thus vertical nodal planes of $A_u^+$ are unstable but the absence of nodes of $A_u^+$ in [001] plane is stable. It seen in Figure 2 c) and d) that $\Psi_{\hat{x}+i\hat{y}}(A_u^-)$ is nodal in [100] plane and is nodeless in [010] plane and that nodal properties of $\Psi_{i\hat{x}-\hat{y}}(A_u^-)$ on both vertical planes are opposite. It is clear that in this case also the nodes in vertical planes may by removed (added) by phase winding $e^{i\varphi}$ and that node in [001] direction is inherent. Going over to the functions of $B_u^\pm$ symmetry we see from Table 2, that adding of a constant phase $i$ in all representation domain results interchange between the case with two nodal vertical planes and with two nodeless vertical planes, but in all cases basal plane is nodal. Thus, considering that there are no experimental lines of nodes of SOP of UTe$_2$ [9, 12] we exclude $B_u^\pm$ cases. The SOP of $A_u^\pm$ with phase winding factor $e^{i\varphi}$ may be chosen nodeless in all planes but in all cases there is a stable point node in [001]- direction. Note, that a group theoretical requirement of $e^{i\varphi}$ phase winding factor for the absence of lines of nodes corresponds to chiral state of UTe$_2$, which was also obtained experimentally [13,14]. In this nonunitary pairing scenario ESP electron pairs of one spin directions break time-reversal symmetry and electrons with opposite spin direction are unpaired. This result is in agreement with experimental data [9,10] also.

In a conclusion, it is shown that the only possible symmetry group corresponding to triplet nonunitary pairs in $D_{2h}$ crystal symmetry is Shubnikov group $m'm'm$ and triplet pair wavefunctions are constructed making use of the space-group approach. It is obtained that Copper pairs of $A_u^\pm$ symmetry may have point nodes only, if phase winding factor $e^{i\varphi}$ in a basis domain of a BZ exists. These theoretical results are in agreement with experimental nonunitary triplet chiral SOP with point nodes in UTe$_2$.

The work was supported by IGIC RAS state assignment.

Table 1. Basis functions of triplet pairs and their nodes for point group $D_{2h}$

| IR | Triplet function | Nodes[1] |
|---|---|---|
| $A_u$ | $\hat{x}(\psi_k - \psi_{C_{2z}k} - \psi_{\sigma_x k} + \psi_{\sigma_y k})$ | [100], $k_y$, $k_z$ |
| $A_u$ | $\hat{y}(\psi_k - \psi_{C_{2z}k} + \psi_{\sigma_x k} - \psi_{\sigma_y k})$ | [010], $k_x$, $k_z$ |
| $A_u$ | $\hat{z}(\psi_k + \psi_{C_{2z}k} + \psi_{\sigma_x k} + \psi_{\sigma_y k})$ | [100], $k_x$, $k_y$ |
| $B_{1u}$ | $\hat{x}(\psi_k - \psi_{C_{2z}k} + \psi_{\sigma_x k} - \psi_{\sigma_y k})$ | [010], $k_x$, $k_z$ |
| $B_{1u}$ | $\hat{y}(\psi_k - \psi_{C_{2z}k} - \psi_{\sigma_x k} + \psi_{\sigma_y k})$ | [100], $k_y$, $k_z$ |
| $B_{1u}$ | $\hat{z}(\psi_k + \psi_{C_{2z}k} - \psi_{\sigma_x k} - \psi_{\sigma_y k})$ | All planes, $k_x$, $k_y$, $k_z$ |
| $B_{2u}$ | $\hat{x}(\psi_k + \psi_{C_{2z}k} + \psi_{\sigma_x k} + \psi_{\sigma_y k})$ | [001], $k_x$, $k_y$ |
| $B_{2u}$ | $\hat{y}(\psi_k + \psi_{C_{2z}k} - \psi_{\sigma_x k} - \psi_{\sigma_y k})$ | All planes, $k_x$, $k_y$, $k_z$ |
| $B_{2u}$ | $\hat{z}(\psi_k - \psi_{C_{2z}k} - \psi_{\sigma_x k} + \psi_{\sigma_y k})$ | [100], $k_y$, $k_z$ |
| $B_{3u}$ | $\hat{x}(\psi_k + \psi_{C_{2z}k} - \psi_{\sigma_x k} - \psi_{\sigma_y k})$ | All planes, $k_x$, $k_y$, $k_z$ |
| $B_{3u}$ | $\hat{y}(\psi_k + \psi_{C_{2z}k} + \psi_{\sigma_x k} + \psi_{\sigma_y k})$ | [001], $k_x$, $k_y$ |
| $B_{3u}$ | $\hat{z}(\psi_k - \psi_{C_{2z}k} + \psi_{\sigma_x k} - \psi_{\sigma_y k})$ | [010], $k_x$, $k_z$ |

[1] Nodal planes are denoted by their normal vectors in square brackets and nodal directions by the axis symbol.

Table 2. Basis functions of triplet ESP pairs for Shubnikov group $m'm'm$ and their nodes. (see footnote of Table 1)

| ICR | Basis function | Nodes |
|---|---|---|
| $A_u^+$ | $(\hat{x}+i\hat{y})(\psi_k - \psi_{C_{2z}k} + \psi_{\sigma_x k} - \psi_{\sigma_y k})$ | [010], $k_x$, $k_z$ |
| $A_u^+$ | $(i\hat{x}-\hat{y})(\psi_k - \psi_{C_{2z}k} - \psi_{\sigma_x k} + \psi_{\sigma_y k})$ | [100], $k_y$, $k_z$ |
| $A_u^-$ | $(\hat{x}+i\hat{y})(\psi_k - \psi_{C_{2z}k} - \psi_{\sigma_x k} + \psi_{\sigma_y k})$ | [100], $k_y$, $k_z$ |
| $A_u^-$ | $(i\hat{x}-\hat{y})(\psi_k - \psi_{C_{2z}k} + \psi_{\sigma_x k} - \psi_{\sigma_y k})$ | [010], $k_x$, $k_z$ |
| $B_u^+$ | $(\hat{x}+i\hat{y})(\psi_k + \psi_{C_{2z}k} + \psi_{\sigma_x k} + \psi_{\sigma_y k})$ | All planes, $k_x$, $k_y$, $k_z$ |
| $B_u^+$ | $(i\hat{x}-\hat{y})(\psi_k + \psi_{C_{2z}k} - \psi_{\sigma_x k} - \psi_{\sigma_y k})$ | [001], $k_x$, $k_y$ |
| $B_u^-$ | $(\hat{x}+i\hat{y})(\psi_k + \psi_{C_{2z}k} - \psi_{\sigma_x k} - \psi_{\sigma_y k})$ | [001], $k_x$, $k_y$ |
| $B_u^-$ | $(i\hat{x}-\hat{y})(\psi_k + \psi_{C_{2z}k} + \psi_{\sigma_x k} + \psi_{\sigma_y k})$ | All planes, $k_x$, $k_y$, $k_z$ |

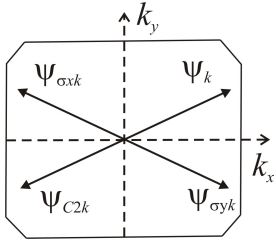

Figure 1. Basis set for Cooper pairs in a BZ for $D_{2h}$ symmetry. Wave vector $k$ corresponds to pair function defined by formula (3). Note that $k$ can have nonzero $k_z$ projection.

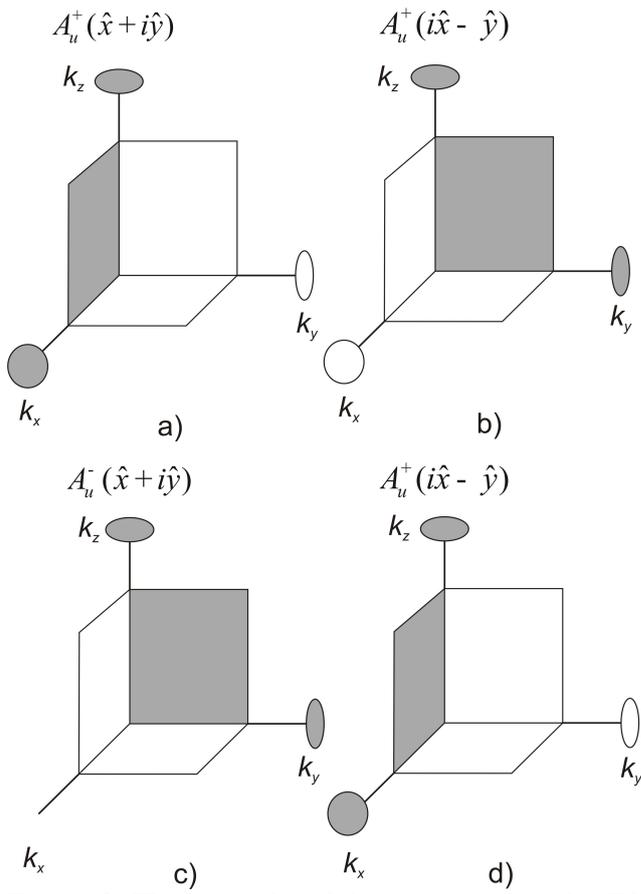

Figure 2. Theoretical nodal structure of triplet ESP Cooper pairs in *k*-space for some ICRs of Shubnikov group $m'm'm$: a) $A_u^+(\hat{x}+i\hat{y})$, b) $A_u^+(i\hat{x}-\hat{y})$, c) $A_u^-(\hat{x}+i\hat{y})$, d) $A_u^-(i\hat{x}-\hat{y})$. Grey (white) squares (circles) correspond to nodal (nodeless) planes (directions) respectively.